\documentstyle[epsf]{l-aa}

\newcommand{\tesc}{t_{\rm esc}}
\newcommand{\tacc}{t_{\rm acc}}
\newcommand{\tbreak}{t_{\rm b}}
\newcommand{\te}{t_{\rm e,0}}
\newcommand{\ta}{t_{\rm a,0}}
\newcommand{\diff}{{\rm d}}
\newcommand{\eqb}{\begin{eqnarray}}
\newcommand{\eqe}{\end{eqnarray}}
\newcommand{\gammamax}{\gamma_{\rm max}}
\newcommand{\numax}{\nu_{\rm max}}
\newcommand{\Gammablob}{\Gamma_{\rm b}}
\newcommand{\betablob}{\beta_{\rm b}}

\newcommand{\tflare}{t_{\rm f}}
\newcommand{\fflare}{\eta_{\rm f}}
\newcommand{\tswitch}{t_{\rm on}}
\newcommand{\xcool}{x_{\rm cool}}

\begin{document}

   \thesaurus{06         
              (03.11.1;  
               16.06.1;  
               19.06.1;  
               19.37.1;  
               19.53.1;  
               19.63.1)} 
   \title{Particle acceleration and synchrotron emission in blazar jets}

\author{J.G. Kirk\inst{1}, F.M. Rieger\inst{2} \& A. Mastichiadis\inst{3}}
\offprints{J.G. Kirk}
\institute{Max-Planck-Institut f\"ur Kernphysik, 
Postfach 10 39 80, D-69029 Heidelberg, Germany
\and
Universit\"atssternwarte, Geismarlandstra{\ss}e 11, D-37083 G\"ottingen, 
Germany
\and
University of Athens, Department of Physics, Panepistimiopolis, 
GR-157 83 Zografos, Greece}
\date{Received 16th.~December 1997, accepted 27th.~January 1998}
\maketitle
\begin{abstract}
We model the acceleration of electrons
at a shock front in a relativistic blazar jet and compute the 
radiation they emit in
a post-shock
region which contains a homogeneous magnetic field. 
The full space, time and momentum dependence of the electron 
distribution is used in this calculation. It is shown that 
the \lq homogeneous\rq\ synchrotron model is recovered, 
provided the downstream speed of 
the plasma away from the shock front is nonrelativistic, and provided
that the light travel times across the face of the shock front is
unimportant. 
By varying the rate at
which particles are picked up by the acceleration process, we calculate  
the time-dependence of the spectra. 
Since the magnetic
field strength is assumed constant within the emission region, each
frequency band can be identified with electrons of a particular energy. We
   find that for a band in which the electrons are accelerated
   rapidly compared to the rate at which they cool, the spectra typically
   harden during phases of rising flux, and soften during phases of falling
   flux, as has been observed in the objects PKS~2155--304 and Mkn~421.                
However, in a frequency band in 
which the timescales are comparable, the reverse behaviour is to be expected. 
We discuss the extent to which observations of both the stationary 
spectrum and the  
spectral variability of the synchrotron component of  
blazar emission can be used to constrain the model.

   \end{abstract}

\section{Introduction}
The nonthermal, highly variable X-ray emission of blazars is usually
interpreted as the synchrotron emission of relativistic electrons accelerated
in a jet which itself moves at relativistic speed towards the observer. In 
this scenario, the higher energy gamma-rays detected from several such sources
arise from the inverse Compton scattering of soft photons by these electrons
(for a review, see Sikora~\cite{sikora94}). Two different types of model of the
emitting region can be found in the literature: the \lq inhomogeneous\rq\
 and the \lq homogeneous\rq\  models. In the inhomogeneous
model, the emitting part of the jet is assumed to be 
cylindrically symmetric with a cross-section which varies with 
distance from the central object. The magnetic field is also taken to vary with
distance from the central object. It is usually 
assumed that relativistic electrons are
\lq injected\rq\ at a shock front which moves through the emission region,
starting close to the central object. These electrons also move outwards with
the jet material, and undergo energy losses due both to synchrotron radiation
and to the adiabatic expansion of the jet (Marscher \& 
Gear~\cite{marschergear85}, Maraschi, Ghisellini \& Celotti 
\cite{maraschietal92}, Marscher \& Travis \cite{marschertravis96}).
The resulting synchrotron spectrum is found by integrating over the emission 
region, within which both the magnetic field and the particle 
distribution is inhomogeneous. 

In the homogeneous model, on the other 
hand, both the magnetic field and the particle distribution function are 
assumed homogeneous throughout the emission region (Inoue \& 
Takahara~\cite{inouetakahara96}, 
Chiaberge \& Ghisellini~\cite{chiabergeghisellini97}, Mastichiadis \& 
Kirk~\cite{mastichiadiskirk97}). The relativistic electrons are injected 
with a specified distribution and are assumed to escape on a timescale 
$\tesc$. After escape, a particle no longer radiates. Although at first sight
somewhat arbitrary, there is a good observational reason to believe that
escape, or, equivalently, sudden energy loss by adiabatic expansion, 
is important: the spectral index of the radio emission of blazars is typically
hard ($\alpha>-0.5$) 
and cannot be produced by particles which have been allowed
to cool completely by synchrotron emission. 
If we adopt a strictly homogeneous model, electrons must 
escape from the emission region and thus be prevented from cooling
completely. 
Such a picture departs from the standard explanation, in which
the flat spectrum is thought to be the result of 
a variation of the self-absorption frequency
within the source in an inhomogeneous model
(the \lq Cosmic Conspiracy\rq\ Marscher~\cite{marscher80}). 
Depending on the magnetic field strength and Doppler boosting factor, 
self-absorption may also become important in 
a homogeneous model. Additional components emitting at low frequency would then
be required.

The homogeneous model with escape corresponds 
approximately to the plausible physical 
situation in which particles are 
accelerated at a shock front, provided there is a 
region of relatively high magnetic field just behind the shock. 
In this case, radiation from this region may
be expected to dominate the observed emission. This requires that on leaving
the region (on a timescale $\tesc$), the particles 
encounter a magnetic field that is so weak, that no significant 
further contribution to the 
emitted spectrum arises, despite the fact that particles accumulate there
over the entire life of the source. 
However, a problem arises in the homogeneous model concerning 
time variability  -- only if the physical
dimensions of the source are such that the light crossing time is short
compared with the synchrotron cooling time 
is it reasonable 
to assume homogeneity. If this condition is not fulfilled, the observed
variability is dominated by the shape and orientation of the source, rather
than the intrinsic cooling and acceleration timescales. Until recently,
models of the rapid variability of blazars have either fulfilled this 
condition or have made specific assumptions about the source geometry
(Mastichiadis \& Kirk~\cite{mastichiadiskirk97}, 
Chiaberge \& Ghisellini~\cite{chiabergeghisellini97}).

In this paper we present a model in which particles are accelerated at a
shock front and cool by synchrotron radiation in the homogeneous magnetic field
behind it. The plasma downstream of the shock front moves relativistically
towards the observer; the shock front is nonrelativistic when 
seen from the rest
frame of the downstream plasma. 
The kinetic equations are solved for the time, space and energy
dependences of the particle distribution function, and the resulting
synchrotron emission is calculated. \lq Escape\rq\ is accounted for by assuming
that the magnetic field strength drops suddenly at a finite distance behind the
shock, so that the radiation from the adiabatically cooled electrons in the
weaker field can be neglected. Thus, this model is homogeneous in the 
sense that the magnetic field does not vary through the emission region, 
but contains an inhomogeneous electron distribution. The
variability is computed by assuming the observer lies in the direction of the 
normal to the shock front. This preferred orientation does not affect the observed 
variability provided the light travel time across the face of the source (i.e.,
over the surface of the shock) is short compared to the synchrotron cooling
time as measured in the rest frame of the plasma. At the highest electron
energies
considered in the application to X-ray blazars, this condition may be 
violated, in which case the computed variability will be smoothed out over
the longer timescale. Whether or not this is expected to occur could be 
decided by 
comparing the predicted emission by inverse compton scattering with 
gamma-ray observations. In principle, observations of the frequency 
dependence of the variation timescale could also resolve the question. 

We consider only the synchrotron radiation of the accelerated particles, 
leaving the more involved computation of the inverse Compton emission to 
future work. This is sufficient for comparison with the observed radio to X-ray 
emission of blazars, provided the energy losses of the electrons are 
not dominated by inverse Compton scattering, which is usually the case,
at least
for BL~Lac-type objects (Comastri~et~al.~\cite{comastrietal97}). 

\section{The electron distribution function}
Consider, then a shock front propagating along a cylindrical jet of 
constant cross-section. Electrons are accelerated at the 
shock front, and subsequently drift away from it in the downstream flow. 
Following Ball \& Kirk~(\cite{ballkirk92}), we treat two spatial zones: 
one around the shock front, in which particles are continuously 
accelerated, and one downstream of it, in which particles emit most of 
the radiation. In each zone, it is assumed that pitch-angle scattering
maintains an almost isotropic particle distribution. 
The number $N(\gamma)\diff \gamma$ of 
particles with Lorentz factor between $\gamma$ and 
$\gamma+\diff\gamma$
in the acceleration 
zone around the shock is governed by the equation
   \eqb
\label{phanomen}
   \frac{\partial N}{\partial t} + \frac{\partial}{\partial \gamma} 
\left[\left(
   {\gamma\over\tacc}  -\beta_{\rm s}\,\gamma^2 \right) N \right] +
   {N\over \tesc}  &=& Q \delta (\gamma - \gamma_0)
   \eqe
(Kirk et al.~\cite{kirketal94}),
   where 
   \eqb
   \beta_{\rm s}&=& 
\frac{4}{3}\frac{\sigma_{\rm T}}{m_{\rm e} c} \left(\frac{B^2}
    {2\,\mu_0} \right) \,.
   \eqe                 
with $\sigma_{\rm T}= 6.65 \cdot 10^{-29} {\rm m}^2$ the 
Thomson cross section. The first term in brackets in 
Eq.~(\ref{phanomen}) describes acceleration at the
rate $\tacc^{-1}$, 
the second describes the rate of energy loss due to synchrotron radiation 
averaged over pitch-angle 
(because of the isotropy of the distribution) 
in a magnetic field $B$ (in Tesla). Particles 
are assumed to escape from this region at an energy independent rate 
$\tesc^{-1}$, and to be picked up (injected) into the acceleration 
process with Lorentz factor $\gamma_0$ at a rate $Q$ particles per 
second. Note that the concept of this \lq acceleration zone\rq\ differs
from the emission region in the usual homogeneous models in two important
respects: a) particles are injected at low energy and continuously accelerated and
b) very little radiation is emitted by a particle whilst in the acceleration
zone (see the discussion below).

Assuming a constant injection rate $Q_0$ after switch-on at time $t=0$, 
and setting $N(\gamma)=0$ for $t\le 0$,
Eq.~(\ref{phanomen}) has the solution
\eqb
\label{dzahl}
N(\gamma,t) &=& a\, \frac{1}{\gamma^2}
\left(\frac{1}{\gamma}-\frac{1}{\gammamax}
\right)^{(\tacc-\tesc)/\tesc} 
\nonumber\\
&&\times\Theta(\gamma-\gamma_0)
\,\Theta(\gamma_1(t)-\gamma) \,,
\eqe
for $\gamma_0<\gamma<\gamma_1(t)$, and $N(\gamma,t)=0$, otherwise. Here
$\Theta(x)$ is the Heaviside step function,
\eqb
\label{con1}
a &=& Q_0\tacc\,\gamma_0^{\tacc/\tesc}\,
\left(1 - \frac{\gamma_0}{\gammamax} 
\right)^{-\tacc/\tesc}
\,,
\eqe
and the upper bound $\gamma_1(t)$ is determined by
\eqb
\gamma_1(t) &=& \left(\frac{1}{\gammamax} + 
\left[\frac{1}{\gamma_0} - 
   \frac{1}{\gammamax}\right]\,{e}^{-t/\tacc}\right)^{-1}\,,
\label{gamma1eq}
\eqe
with $\gammamax=(\beta_{\rm s}\tacc)^{-1}$. The quantities $\tacc$ and $\tesc$
have been assumed independent of energy in this solution, the more general
solution allowing for energy dependence is given in the Appendix.

\begin{figure}
\epsfxsize 8 cm
\epsffile{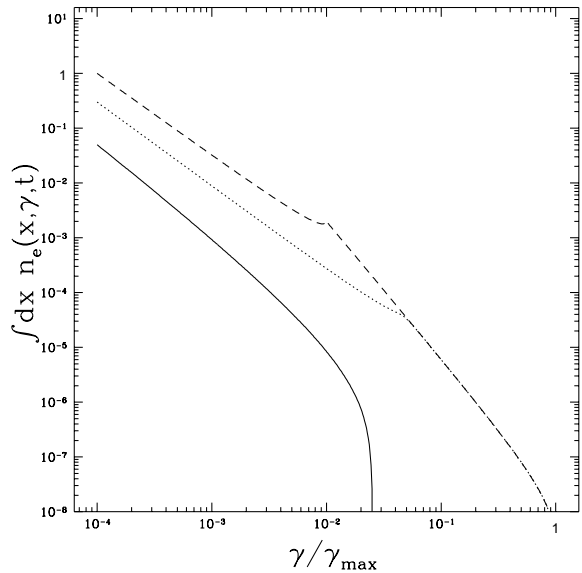}
\caption{
\label{elecdist}
The electron distribution integrated over the source, as given by 
Eq.~(\protect\ref{intdensity}). The three curves correspond to the
times $5\protect\tacc$ (solid line), $30\protect\tacc$ (dotted line) and 
$500\protect\tacc$, (dashed line). At larger times the distribution does not 
evolve appreciably. For this plot, 
$\protect\tesc=2\protect\tacc$ (i.e., $s=3.5$) and $\protect\tbreak=100\tacc$.
At $t=5\tacc$, no particles have had time to cool, since the cooling time
at the maximum Lorentz factor of
$\protect\gamma\protect\approx0.025\protect\gammamax$ is approximately 
$40\tacc$. At times $t>500\protect\tacc$ all particles with 
$\protect\gamma>\protect\gammamax\protect\tacc/\protect\tbreak$ 
cool, whereas those of lower
$\protect\gamma$ leave the source without significant loss of energy.
}
\end{figure}

In the model developed by Ball \& Kirk~(\cite{ballkirk92}), accelerated 
particles escape into the downstream plasma, where they radiate. 
We can formulate the kinetic equation obeyed by the 
density of particles in the radiation zone most compactly using a 
coordinate system at rest in the radiating plasma. The shock front then 
provides a moving source of electrons, which subsequently suffer energy 
losses, but are assumed not to be transported in space. The kinetic 
equation governing the differential 
density $\diff n(x,\gamma,t)$ of particles in 
the range $\diff x$, $\diff \gamma$ is then
\eqb
\label{trans}
   \frac{\partial n}{\partial t} -
\frac{\partial}{\partial \gamma}(\beta_{\rm s}\,\gamma^2\,n) &=&
{N(\gamma,t)\over\tesc}\delta(x-x_{\rm s}(t)) 
\eqe
where $x_{\rm s}(t)$ is the position of the shock front at time $t$.
For a shock which starts to accelerate (and therefore \lq 
inject\rq) particles at time $t=0$ and position $x=0$ and moves
at constant speed $u_{\rm s}$, the solution of Eq.~(\ref{trans}) 
for $\gamma>\gamma_0$ is 
\eqb
n(x,\gamma,t)&=&
{a\over u_{\rm s}\tesc\gamma^2}
\nonumber\\
&&\left[\frac{1}{\gamma}-\beta_{\rm s} \left(t -
  \frac{x}{u_{\rm s}}\right) - 
\frac{1}{\gammamax}\right]^{(\tacc-\tesc)/\tesc}
\nonumber\\
&&
\Theta \left[\gamma_1(x/u_{\rm s})- 
   (1/\gamma - \beta_{\rm s}t+ \beta_{\rm s}x/u_{\rm s})^{-1}\,\right]
\,,
\label{fulldist}
\eqe
where $\gamma_1(t)$ is given by Eq.~(\ref{gamma1eq}).
To obtain the synchrotron emissivity as a function of position, time and
frequency we convolve the density $n$ with the
synchrotron Green's function $P(\nu,\gamma)$. A convenient approximation to 
this function is given by Melrose~(\cite{melrose80}):
\eqb
P(\nu,\gamma)&=&a_{\rm s} z^{0.3}\exp({-z})
\label{synchgreen}
\eqe
where $a_{\rm s}=\sqrt{3}e^2\Omega\sin\theta/(2\pi c)$ 
is a constant and 
$z\equiv 4\pi\nu/(3\Omega\sin\theta\gamma^2)$, with $\Omega=eB/m$ the electron
gyro frequency and $\theta$ the angle between the magnetic field 
direction and the
line of sight.
At a point  $x=X$ ($>u_{\rm s}t$)
on the symmetry axis 
of the source at time $t$
the specific intensity of radiation in the $\vec{x}$ direction 
depends on the retarded time ${\bar t}=t-X/c$ and is given by
\eqb
\label{jetframe}
I_0(\nu,{\bar t})\,=\,\int\diff\gamma P(\nu,\gamma)\int\diff x
\,n(x,\gamma,{\bar t}+x/c)
\eqe
and the integrated particle density can readily be evaluated:
\eqb
\lefteqn{
\int \diff x \,n(x,\gamma,{\bar t}+x/c)
\,=}
\nonumber\\
&&{a\over (1-u_{\rm s}/c)
\gammamax^{(\tacc+\tesc)/\tesc}}
\left({\gammamax\over\gamma}\right)^2
\nonumber\\
&&\left\lbrace
\left[\frac{\gammamax}{\gamma}-{{\bar t}\over\tacc} +
  \frac{x(1-u_{\rm s}/c)}{u_{\rm s}\tacc} - 1
\right]^{\tacc/\tesc}\right\rbrace^{x_1({\bar t})}_{x_0(\gamma,{\bar t})}
\,.
\label{intdensity}
\eqe 
The limits of the spatial integration are given by the 
retarded position of the shock front
\eqb
x_1({\bar t}) &=& 
{u_{\rm s}{\bar t}\over 1-(u_{\rm s}/c)}\,.
\eqe
and the retarded position $x_0(\gamma,{\bar t})$ 
of the point furthest from the shock front at which particles
have Lorentz factor $\gamma$ at time ${\bar t}+x_0/c$. This is given either 
by the solution $\xcool(\gamma,\bar t)$ of the transcendental equation
\eqb
\lefteqn{
\left[{\gammamax\over\gamma}-{{\bar t}+\xcool/c\over\tacc}+
{\xcool\over u_{\rm s}\tacc}\right]\,=}
\nonumber\\
&&\,1+\left({\gammamax\over\gamma_0} -1\right)
\exp\left[{-\xcool/(u_{\rm s}\tacc)}\right]
\eqe
or by the assumed maximum spatial extent of the emission region (i.e.,
the point at which the magnetic field declines substantially). 
Denoting this distance by $L$, we have 
\eqb
x_0(\gamma,\bar t)&=& {\rm Max}\left[\xcool(\gamma,\bar t),x_1(\bar t)
-L\right]
\label{coollimit}
\eqe
This is most conveniently expressed in terms of the (retarded) time 
$\tbreak$ for the plasma to traverse the
emitting region, as measured in the plasma rest frame:
\eqb
\tbreak&=&{ (1-u_{\rm s}/c)L\over u_{\rm s}}
\eqe
Equation (\ref{intdensity}) gives the integrated particle density for
times greater than the \lq switch-on\rq\ time:
\eqb
{\bar t}&>&\tswitch\,=\,\tacc\left(1-{u_{\rm s}\over c}\right)
\log\left[{(\gammamax/\gamma_0)-1\over
(\gammamax/\gamma) - 1}\right]
\eqe
before which it vanishes.
The resulting electron spectrum, integrated over the source is depicted in 
Fig.~\ref{elecdist}. A characteristic break in the spectral slope appears
at a particular Lorentz factor $\gamma_{\rm br}(t)$ which, at 
any given time, separates those electrons which cool within the source
($\gamma>\gamma_{\rm break}$) from those which do not cool within the source
($\gamma<\gamma_{\rm break}$). 
At large times, all electrons with
$\gamma<\gamma_{\rm break}(\infty)=\gammamax/(1+\tbreak/\tacc)$ leave the 
source before cooling, and the integrated electron density becomes  
time-independent.

A question which remains open in this approach is the 
synchrotron radiation emitted by a particle whilst in the acceleration
region. If the magnetic field ahead of the shock were the same as that
behind the shock, the total emission could easily be computed using 
Eq.~(\ref{dzahl}).
For $t>\tswitch$, we would have an extra contribution to the flux:
\eqb
\lefteqn{I_{\rm s}(\nu,\bar{t})\,=\,\int\diff\gamma P(\nu,\gamma)
\int\diff x N(\gamma,{\bar t})\delta[x-u_{\rm s}({\bar t}+x/c)]}
\nonumber\\
&&\,=\,{a\over (1-u_{\rm s}/c)
\gammamax^{(\tacc+\tesc)/\tesc}}
\int\diff\gamma P(\nu,\gamma)\left({\gammamax\over\gamma}\right)^2
\nonumber\\
&&\left(\frac{\gammamax}{\gamma}-1\right)^{(\tacc-\tesc)/\tesc}
\,.
\label{shockemission}
\eqe
and the total emission $I_1(\nu,\bar t)$ would be given by
\eqb
I_1(\nu,\bar t)&=& I_0(\nu,\bar t)+I_{\rm s}(\nu,\bar t)
\eqe

However, for oblique shocks, the magnetic field strength is expected to
increase upon compression at the shock. Particles undergoing acceleration spend
part of the time in the upstream and part in the downstream plasma, so that it
is not clear how to evaluate their synchrotron emission, although 
Eq.~(\ref{shockemission}) certainly gives an upper limit. At 
oblique shocks, reflection at the 
front itself
is thought to be more important than diffusion in the downstream zone, 
(Kirk \& Heavens~\cite{kirkheavens89}), 
so that accelerating particles spend all their time 
upstream. In this case, it seems reasonable to neglect the
emission from the acceleration zone completely, which is the approach
adopted here. An improved treatment of this point demands a full time-dependent
solution of the diffusion advection equation, which 
requires considerable numerical effort (Fritz \& Webb~\cite{fritzwebb90}).

All quantities calculated so far in this section refer to 
the frame in which the radiating plasma is at rest
(the jet frame). For 
application to blazars, they must be transformed into the observer's 
frame. Assuming the observer to lie in the direction of motion of the 
plasma, and denoting the plasma bulk 
velocity by $c\betablob$, the relevant transformations are
\eqb
\label{stretch}
X&=& \Gammablob (D- \betablob c\,t_{\rm obs}   )
\\
\label{timedilation}
t&=&\Gammablob( t_{\rm obs}  - \betablob D/c)
\\
\nu_{\rm obs}&=&\Gammablob(1+\betablob) \nu\approx 2\Gammablob\nu
\\
\label{dopplerboost}
I_{\rm obs}({\nu_{\rm obs}},t_{\rm obs})&=&
(\nu_{\rm obs}/\nu)^3 I(\nu,{\bar t})
\,\approx\,8\Gammablob^3 I(\nu,{\bar t})
\eqe
where $\Gammablob$ ($\gg 1$) 
is the Lorentz factor of the approaching jet plasma 
and $D$ the position of the detector with 
respect to the position of the shock at $t=0$, as measured in the 
observer's reference frame. 

Several simple qualitative results follow from these expressions. Close to 
the maximum emitted frequency, the timescale on which the intensity varies
in the frame of the plasma is roughly the switch-on time 
$\sim (1-u_{\rm s}/c)\tacc$. According to Eqs.~(\ref{stretch}) and 
(\ref{timedilation}), ${\bar t}=(1+\betablob)\Gammablob(t_{\rm obs}-D/c)$ 
Thus, the observed timescale is shorter than the intrinsic $\tacc$
by a factor $\Gammablob(1+\betablob)/(1-u_{\rm s}/c)$. 
Doppler boosting of the flux, is independent of the 
shock speed, and is given simply by Eq.~(\ref{dopplerboost}).

\section{Application to the synchrotron spectra of blazars}
\begin{figure}
\epsfxsize 8 cm
\epsffile{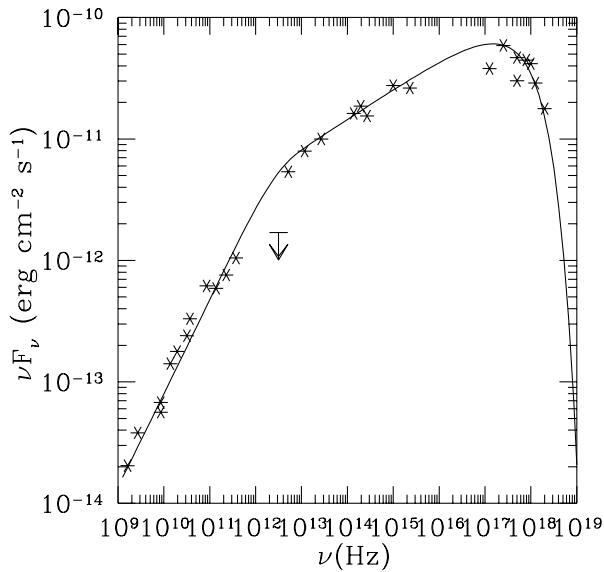}
\caption{
\label{mkn501}
The radio -- X-ray spectrum of the object Mkn~501 (data taken 
from the collation of 
Catanese et al.~\protect\cite{cataneseetal97}) together with the 
stationary synchrotron
emission from a single homogeneous source
}
\end{figure}

The simple model described above provides a remarkably good fit to the
radio to X-ray spectra of several blazars. As an example, we show in 
Fig.~\ref{mkn501} observations of the object Mkn~501.
The gamma-ray emission of this object,
which has been the subject of much recent interest
(e.g., Bradbury et al~\cite{bradburyetal97}), is not included in this
figure, since it is not thought to arise as synchrotron radiation.
For the X-ray emission we display the archival data selected by
Catanese et al.~(\cite{cataneseetal97}) and not the data taken during the TeV
flare in April~1997. 

The stationary emission ($t\rightarrow\infty$) found from Eq.~(\ref{jetframe})
by a single numerical quadrature over the synchrotron Green's function is also
shown in Fig.~\ref{mkn501}. The form of the spectrum is very close to that 
given by Meisenheimer \& Heavens~(\cite{meisenheimerheavens87}), who
used an analytic solution to the stationary diffusion/advection equation,
including synchrotron losses. Four free parameters are used to produce this 
fit: 
\begin{enumerate}
\item
the low frequency spectral index $\alpha=-0.25$, which corresponds to
taking $\tacc=\tesc/2$ 
\item
the characteristic synchrotron frequency emitted by an
electron 
of the maximum Lorentz factor 
as seen in the observers frame 
(taken to be $1.3\times10^{18}\,$Hz) 
\item
the spatial extent of the emitting region, which determines the 
position of the spectral break at roughly $5\times10^{12}\,$Hz,
corresponding to $\tbreak=700\tacc$ 
\item
the absolute flux level.
\end{enumerate}
  
Since we restrict our model to the synchrotron emission of the accelerated
particles, it is not possible independently 
to constrain quantities such as the Doppler
boosting factor, or the magnetic field. These can, however, be found using a
model for the gamma-ray emission, for example the synchrotron self-compton
model (Mastichiadis \& Kirk~\cite{mastichiadiskirk97}). 
Similarly, the frequency below which synchrotron self-absorption modifies the
optically thin spectrum is not constrained within the current picture.
Nevertheless, our model
of the synchrotron emission 
makes predictions concerning the spectral variability in each of the three
characteristic frequency ranges which can be identified in
Fig.~\ref{mkn501}. These ranges are generic features of any synchrotron model,
so that the predicted variability can easily 
be applied to the synchrotron emission of
other blazars. They are a) the low frequency region, where the particles have
not had time to cool before leaving the source (this is the region with
$\alpha=-0.25$ in Fig.~\ref{mkn501}, below the break at
$5\times10^{12}\,$Hz) b) the region between the break and the maximum flux,
where the particles have had time to cool, but where the cooling rate is always
much slower than the acceleration rate and the spectrum is close to 
$\alpha=-0.75$, and c) the region around and
above the flux maximum at roughly $10^{17}\,$Hz, where the acceleration rate is
comparable to the cooling rate.

Variability or flaring behaviour can arise for a number of reasons.
When the shock front overruns a region in the jet in which the local plasma
density is enhanced, the number of particles picked up and injected into the
acceleration process might be expected to increase. In addition, if the density
change is associated with a change in the magnetic field strength, the
acceleration timescale might also change, and, hence, the maximum frequency
of the emitted synchrotron radiation. Considering the case in which the
acceleration timescale remains constant, it is a 
simple matter to compute the emission, since
Eq.~(\ref{trans}) is linear.
An increase of the injection rate by a factor $1+\fflare$ for a time
$\tflare$ is found by setting
\eqb
Q(t)&=&Q_0\qquad {\rm for }\ t<0\ {\rm and}\ t>\tflare
\\
Q(t)&=&(1+\fflare) Q_0 \qquad {\rm for}\ 0<t<\tflare
\label{flareeq}
\eqe
We then have
\eqb
\lefteqn{I(\nu,{\bar t})= I_1(\nu,\infty)+}
\nonumber\\
&&\fflare\left[
I_1(\nu,{\bar t}) - I_1(\nu,{\bar t}-(1-u_{\rm s}/c)\tflare)\right]
\label{flareexpr}
\eqe

\begin{figure}
\epsfxsize 8 cm
\epsffile{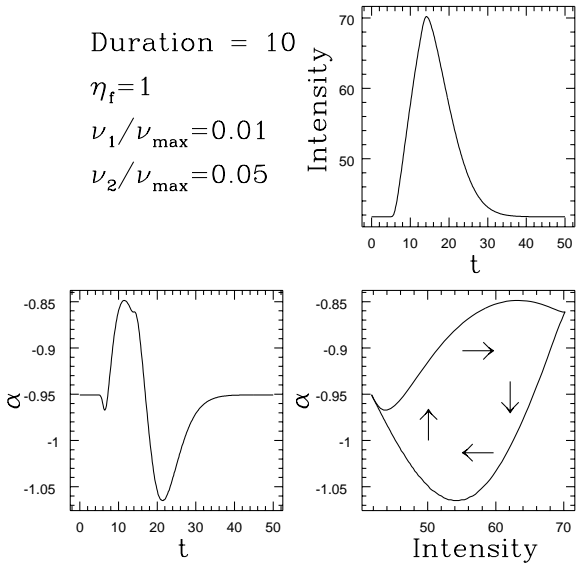}
\caption{\protect\label{loop1}
The intensity and spectral index during the flare
described by Eq.~(\protect\ref{flareeq}), as a function of time at 
low frequency. The loop in the $\protect\alpha$ vs.\ intensity plot
is followed in the clockwise direction.}
\end{figure}

Using $\fflare =1$, $\tflare=10\tacc$ and $u_{\rm s}=c/10$, 
we show the resulting emission at
a frequency $\nu=\numax/100$ in 
Fig.~\ref{loop1}. In the case of Mkn~501, this corresponds to a frequency of
about $10^{16}\,$Hz, which lies towards the high frequency part of 
region b),
between the infra-red and X-ray
regions, where the spectral index is close to $\alpha=-0.75$. 
Also shown in this figure is the temporal behaviour of the 
spectral index, as determined from the ratio of the fluxes at $0.01\numax$ and 
$0.05\numax$, through the flare. When plotted against the flux at the lower
frequency, the spectral index exhibits a characteristic loop-like pattern,
which is tracked in the clockwise sense by the system.  
This type of pattern is well-known and has 
been observed at different wavelengths in several sources e.g., 
OJ~287 (Gear et al.~\cite{gearetal86}), 
PKS~2155--304 (Sembay et al.~\cite{sembayetal93}) and  
Mkn~421 (Takahashi et al.~\cite{takahashietal96}). 
It arises
whenever the spectral slope is controlled by synchrotron cooling, (or,
in fact, any cooling process which is faster at higher energy) so that
information about changes in the injection propagates from high to low 
energies (Tashiro et al.~\cite{tashiroetal95}).

\begin{figure}
\epsfxsize 8 cm
\epsffile{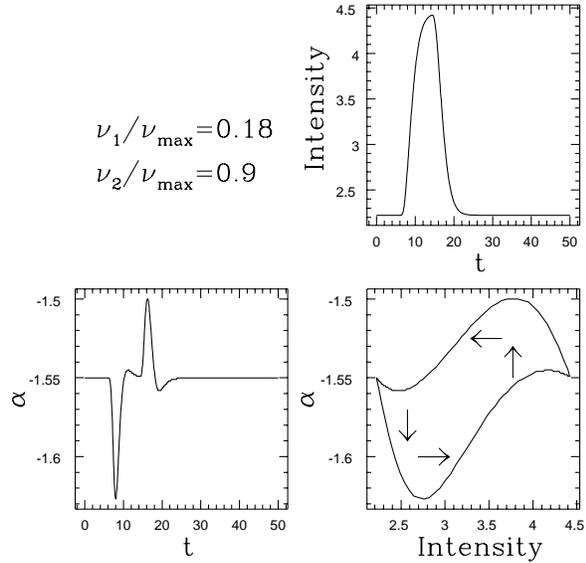}
\caption{\protect\label{loop2}
The intensity and spectral index in the same flare as in 
Fig.~\protect\ref{loop1}
but at 
high frequency. The loop in the $\protect\alpha$ vs.\ intensity plot
is followed in the anticlockwise direction.}
\end{figure}

If the system is observed closer to the maximum frequency, where the cooling
and acceleration times are equal, the picture changes. Here information about
the occurrence of a flare propagates from lower to higher energy, as particles
are gradually accelerated into the radiating window. Such behaviour is depicted
in Fig.~\ref{loop2}, where the same flare is shown at frequencies which are 
an order of magnitude higher than in Fig.~\ref{loop1}. 
In the case of Mkn~501, the frequency range is close to $10^{18}\,$Hz.
This time the loop is traced anticlockwise.
Such behaviour, although not as common, 
has also occasionally been observed, for example in the case of 
PKS~2155--304 (Sembay et al.~\cite{sembayetal93}).

Finally, in the region of the spectrum below the break, 
where the spectral index in the case of Mkn~501 is close to
$\alpha=-0.25$, the emission is determined by the finite size of the
source region, as expressed by $\tbreak$. Here, the flare shown in
Figs.~\ref{loop1} and \ref{loop2} has only a very small effect on the observed
flux, since the duration $\tflare=10\tacc$ has been chosen to be much smaller
than the time $\tbreak$ taken to fill up the emitting region with 
radiating particles. However, even in the case
of a larger flare, or one of longer
duration, no variation of the spectral index is to be expected through the
flare at frequencies below the break, because the time taken to fill the
effective emitting region is independent of frequency.

\begin{figure}
\epsfxsize 8 cm
\epsffile{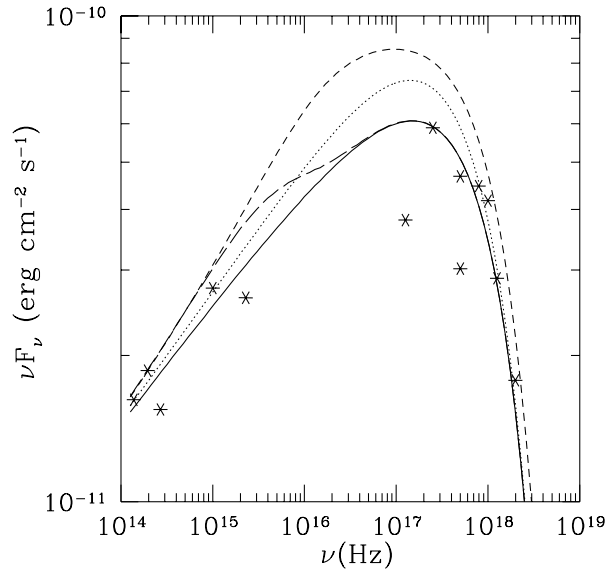}
\caption{\protect\label{timedep}
The spectrum at times $t=0$ (solid line), 10 (dotted), 20 (short-dashed)
and $30\protect\tacc$ (long-dashed) during the flare shown in 
Figs.~\protect\ref{loop1} and \protect\ref{loop2}. Data points are taken from 
Catanese~et~al.~(\protect\cite{cataneseetal97}).
}
\end{figure}

This effect of smaller changes in the spectral slope at lower frequencies 
is also evident from Fig.~\ref{timedep}, 
where the spectrum is shown at times $t=0$
(i.e., the stationary emission as shown in Fig.~\ref{mkn501}), $t=10\tacc$, 
$t=20\tacc$,
and $t=30\tacc$. Here it can be seen that the rise in emission is rapid at all
frequencies, and the subsequent fall sets in as a wave which propagates
downwards in frequency.

The results presented in Figs.~\ref{loop1} and \ref{loop2} 
are computed for $u_{\rm s}=1/10$, so that the effects of the finite 
light travel time between the front and back of the source are 
negligible. In fact, in the limit $u_{\rm s}\rightarrow 0$, 
Eqs.~(\ref{fulldist}) to (\ref{flareexpr}) reduce to the equations for a 
homogeneous source, provided the limit is taken keeping $\tbreak$ 
finite, i.e., allowing the maximum spatial extent of the source 
$u_{\rm s}\tbreak$ to vanish.

\section{Discussion}
It is generally accepted that the emission mechanism responsible for the 
radio to optical emission of blazars is the synchrotron process. In objects 
similar to Mkn~501, which is discussed in Sect.~3, the synchrotron emission 
extends up to X-ray energies. It has recently emerged that a model in 
which the radiating relativistic electrons reside in a region of uniform 
magnetic field can provide a reasonable fit to observations not only of 
the synchrotron component, but also, via the inverse compton scattering 
of the synchrotron photons, 
to gamma-ray observations in the GeV to TeV range.
In this paper, we have introduced a detailed model of the emission 
region
including the processes of acceleration and synchrotron cooling. 
This should ultimately help 
us to assess how accurately a homogeneous region can 
reproduce the observed synchrotron emission. In combination with a 
computation of the inverse compton emission, the physical conditions and 
the geometry of the emission zone can be constrained.

The model used to describe acceleration is similar to that used by Ball
\& Kirk~(\cite{ballkirk92}) to describe the nonthermal radio 
emission of SN1987A. It involves a phenomenological division of the 
particle population into two groups: particles which are undergoing both 
acceleration and cooling and particles which merely cool. As a result, 
distributions are found which extend to a maximum value of the Lorentz 
factor, and then cut-off abruptly. The synchrotron emission from such a 
distribution fits well with that from the source we have discussed in 
Sect.~3 -- Mkn~501. However, the form of the spectrum, especially in the hard
X-ray region, depends on the shape of the cut-off of the electron spectrum.
In our model, the sharpest possible turn-over is produced. Inhomogeneities
within the source or refinements of the acceleration model 
result in a broader turn-over. This indicates 
that in the case of Mkn~501 a homogeneous model of the region in which
synchrotron X-rays are emitted is a good approximation. It also confirms that
a detailed model of the cut-off -- which would necessarily involve additional
parameters -- is unnecessary. In other blazars e.g., Mkn~421 as well as in
other synchrotron emitting objects e.g., SNR1006 (Mastichiadis \& 
de~Jager~\cite{mastichiadisdejager96}), a broader cut-off is indicated by the 
observations. Our current computations suggest that if the physical situation
is similar to that in Mkn~501, the broadening of the cut-off
should be attributed to inhomogeneities within the source rather than an
intrinsic property of the acceleration mechanism.

The variability predicted for a homogeneous source depends on the frequency of
observation and the parameters of the acceleration mechanism -- for relatively
low frequency radiation, a characteristic pattern is produced, as pointed out
by several groups. We have shown that closer to the maximum emitted frequency,
this pattern should change. However, the observed variability depends not only
on the intrinsic time-dependence, but also on the smoothing caused by light
travel time delays across the source. We have described in detail the situation
when the observer is positioned exactly on the axis of the source so that the
emitting plasma moves directly towards the observer. This assumption is not as
restrictive as it might at first sight appear, since the emission from a source
in relativistic motion with Lorentz factor $\Gamma$ 
is significantly boosted when viewed from a direction
which makes an angle of less than $\sim1/\Gamma$ with the velocity. Intrinsic
time variations within the source then appear shorter 
by a factor $\Gamma$ (see
Sect.~2). These are also smoothed out 
on the timescale $R/(c\Gamma)$, where $R$ is a
typical dimension of the source perpendicular to the line of sight, provided
that the angle between the source velocity and line of sight is of the order of
$1/\Gamma$. It is a good approximation to neglect this smoothing if the
intrinsic synchrotron cooling time, measured in the rest frame of the emitting
plasma, is longer than the light travel time $R/c$. 
The same restriction applies to the size of the source along the 
direction to the observer. Here, however, the intrinsic variations always 
dominate if $u_{\rm s}\ll c$. The general formulae given in Sect.~2 
are valid also for relativistic $u_{\rm s}$, but in the examples 
predicted, we have restricted ourselves to $u_{\rm s}=c/10$.  
Whether or not intrinsic variability dominates over light travel time 
effects depends on the frequency of observation. 
For Mkn~501, for example, we have found that the maximum timescale
over which particles cool is determined by the position of the spectral break
and is roughly $700$ times the synchrotron cooling time 
at the maximum emitted frequency. Thus, depending on $R$, there may exist a
critical frequency 
above which variations are smoothed out by light travel time
effects, but below which the results of Sect.~3 are valid. A detailed model of
the inverse Compton emission is needed to estimate this frequency.

\begin{acknowledgements}
We acknowledge support for this work from the 
Deutsche Forschungsgemeinschaft under SFB~328 (A.M.) and DFG Ma~1545/2-1
(F.R.).
\end{acknowledgements}

\appendix
\section{Energy dependent acceleration and escape}
In general, one might expect the acceleration time $\tacc$ and escape
time $\tesc$ introduced in Eq.~(\ref{phanomen}) to be functions of particle
energy. In this case the solution Eq.~(\ref{dzahl}) is modified. Defining
\eqb
f(\gamma)&=&
{\gamma\over\tacc(\gamma)} - \beta_{\rm s}\gamma^2
\eqe
which is positive in the range of interest, one can write
\eqb
\label{phanomen2}
   \frac{\partial N}{\partial t} + \frac{\partial}{\partial \gamma} 
\left[f(\gamma) N \right] +
   {N\over \tesc(\gamma)}  &=& Q_0\Theta(t) \delta (\gamma - \gamma_0)
\enspace.
\eqe
The solution of this equation subject to the boundary condition 
$N(\gamma,0)=0$ is easily found using Laplace transforms: 
\eqb
N(\gamma,t)&=&{Q_0\over f(\gamma)}
\Theta\left[t-\tau(\gamma)\right]
{\rm exp}\left[ - \int_{\gamma_0}^{\gamma}
{\diff \gamma'\over \tesc(\gamma')f(\gamma')}\right]
\eqe
where
\eqb
\tau(\gamma)&=&\int_{\gamma_0}^\gamma{\diff\gamma'\over f(\gamma')}
\eqe
In addition to the straightforward case $\tacc=\,$constant, $\tesc=\,$constant
dealt with above, it is also interesting to consider the case in which both
of these quantities are linearly proportional to $\gamma$. This would arise in 
modelling diffusive acceleration with a 
\lq gyro-Bohm\rq\ spatial diffusion coefficient (e.g., Kirk et
al.~\cite{kirketal94}).
The solution is 
\eqb
N(\gamma,t)&=&\Theta\left[t-\tau(\gamma)\right]
\nonumber\\
&&{Q_0\ta\gammamax^2\over(\gammamax^2-\gamma^2)}
\left[{\gamma(\gammamax^2-\gamma_0^2)\over
\gamma_0(\gammamax^2-\gamma^2)}\right]^{-\ta/\te}
\eqe
with
\eqb
\tau(\gamma)&=&{\ta\over2}\log\left[(\gammamax+\gamma)(\gammamax-\gamma_0)\over
(\gammamax-\gamma)(\gammamax+\gamma_0)\right]
\eqe
and where we have written
\eqb
\tacc(\gamma)=\gamma\ta &\qquad&\tesc(\gamma)=\gamma\te
\eqe
Note that in this case the power-law index $s$, defined by
$N\propto\gamma^{-s+2}$ is 
related to the timescales by $s=2+\ta/\te$, whereas in the case of energy
independent acceleration and escape $s=3+\tacc/\tesc$.

\end{document}